\documentclass[prd,aps,twocolumn,nofootinbib,preprintnumbers,superscriptaddress,preprintnumbers,balancelastpage,longbibliography,10pt]{revtex4-2}
\usepackage[english]{babel}

\usepackage{amsmath,mathtools,physics,xfrac}
\usepackage{graphicx}
\usepackage{afterpage}
\usepackage{subfigure}
\usepackage{rotating}
\usepackage{multirow}
\usepackage{tabularx}
\usepackage{booktabs}
\usepackage{fancyhdr}
\usepackage[utf8]{inputenc}
\usepackage{theorem}
\usepackage{moreverb}
\usepackage{euscript}
\usepackage{psfrag}
\usepackage{slashed}
\usepackage{mathtools}
\usepackage{makecell}
\usepackage{adjustbox}
\usepackage{dcolumn}
\usepackage{bm}
\usepackage[dvipsnames]{xcolor}
\usepackage{graphics}

\usepackage{chngcntr}
\usepackage{aas_macros}
\usepackage{tabularx}
\usepackage{color,xcolor}

\usepackage{hyperref}

\newcommand{\blfootnote}[1]{%
  \begingroup
  \renewcommand{\thefootnote}{}\footnote{#1}%
  \addtocounter{footnote}{-1}%
  \endgroup
}

\usepackage[normalem]{ulem}
\hypersetup{
     colorlinks   = true,
     citecolor    = Green,
     urlcolor     = Green,
     linkcolor    = Green
}

\definecolor{darkblue}{rgb}{0.0,0.0,0.75}
\definecolor{darkred}{rgb}{0.6,0.0,0}
\definecolor{darkgreen}{rgb}{0.0,0.6,0.}

\counterwithout{equation}{section}

\usepackage{tikz}

\definecolor{lime}{HTML}{A6CE39}
\DeclareRobustCommand{\orcidicon}{\hspace{-1mm}
	\begin{tikzpicture}
		\draw[lime, fill=lime] (0,0) 
		circle [radius=0.16] 
		node[white] {{\fontfamily{qag}\selectfont \tiny \,ID}};
		\draw[white, fill=white] (-0.0525,0.095) 
		circle [radius=0.007];
	\end{tikzpicture}
	\hspace{-3mm}
}

\foreach \x in {A, ..., Z}{\expandafter\xdef\csname orcid\x\endcsname{\noexpand\href{https://orcid.org/\csname orcidauthor\x\endcsname}
		{\noexpand\orcidicon}}
}


\begin{document}


\title{\boldmath Hunting Sterile Neutrino Dark Matter in the MeV Gap}

\author{Shivam Gola\,\orcidA{}}
\email{shivamg.sk@iitb.ac.in}
\affiliation{Department of Physics, Indian Institute of Technology Bombay, Mumbai 400076, India}

\author{Akash Kumar Saha\orcidB{}}
\email{akashks@iisc.ac.in}
\affiliation{Centre for High Energy Physics, Indian Institute of Science, C.\,V.\,Raman Avenue, Bengaluru 560012, India}

\author{Manibrata Sen\orcidC{}}
\email{manibrata@iitb.ac.in}
\affiliation{Department of Physics, Indian Institute of Technology Bombay, Mumbai 400076, India}

	
	
	\begin{abstract}
We investigate the sensitivities of upcoming MeV gamma-ray telescopes to sterile neutrino dark matter in the mass range $(0.2-100)\,{\rm MeV}$. Sterile neutrinos in this regime can produce observable photon signals through radiative two-body decays and three-body decays with final-state radiation. We perform a Fisher forecasting analysis incorporating realistic astrophysical background modeling and detector response to derive projected constraints on the sterile neutrino decay rate. We find that future MeV instruments can improve existing limits by several orders of magnitude across a wide region of parameter space. Our results highlight the discovery potential of next-generation MeV telescopes in probing sterile neutrino dark matter.

	\end{abstract}
	
	\maketitle

\blfootnote{SG and  AKS contributed equally to this work.}

\section{\label{sec:level1}Introduction}
The existence of dark matter (DM) and the presence of a non-zero neutrino masses remain two of the most important motivations for physics beyond the Standard Model (BSM)\,\cite{Cirelli:2024ssz, Slatyer:2017sev,Strigari:2012acq}. An extra species of inert neutrinos, aptly termed sterile neutrinos, is one of the few promising candidates that can resolve both of these issues and additionally provide key insights to the origin of the baryon asymmetry in the Universe\,\cite{Kusenko:2009up,Abazajian:2017tcc,Drewes:2016upu,Dasgupta:2021ies, Asaka:2005an, Boyarsky:2006jm, Asaka:2006nq, Canetti:2012kh}. While the SM contains only three species of active neutrinos, cosmological observations allow the presence of one or more sterile neutrinos, depending on their masses and mixings. Sterile neutrino searches from various ground and space-based observatories have provided stringent bounds on its mass and mixing with active neutrinos.

Sterile neutrino DM can be produced in the early Universe through a variety of mechanisms. In the keV mass range, the most widely studied scenarios include non-resonant production via active–sterile neutrino oscillations (the Dodelson–Widrow mechanism)\,\cite{Dodelson:1993je} and resonant production in the presence of a lepton asymmetry (the Shi–Fuller mechanism)\,\cite{Shi:1998km}. While the former is now strongly constrained by a combination of X-ray and structure formation bounds, the latter remains viable in restricted regions of parameter space. Beyond these minimal frameworks, a broad class of models involving additional interactions or fields can produce sterile neutrinos through decay or freeze-in processes, thereby relaxing cosmological constraints and modifying their phase space distribution\,\cite{Shaposhnikov:2006xi,Petraki:2007gq,Merle:2013wta,Shuve:2014doa,Shakya:2015xnx,Shakya:2016oxf,Hansen:2017rxr,Patwardhan:2015kga,Herms:2018ajr,Bezrukov:2017ike,Bezrukov:2018wvd,Bezrukov:2019mak,Kelly:2020pcy,Kelly:2020aks,Benso:2021hhh}. Such scenarios naturally extend to heavier sterile neutrino masses in the MeV range, where production can occur via the decay of heavier particles, scalar portals, or new gauge interactions. This motivates a systematic exploration of keV-MeV scale sterile neutrino DM using next-generation detectors.
 
\begin{figure}[t]
\centering
\includegraphics[width=\columnwidth]{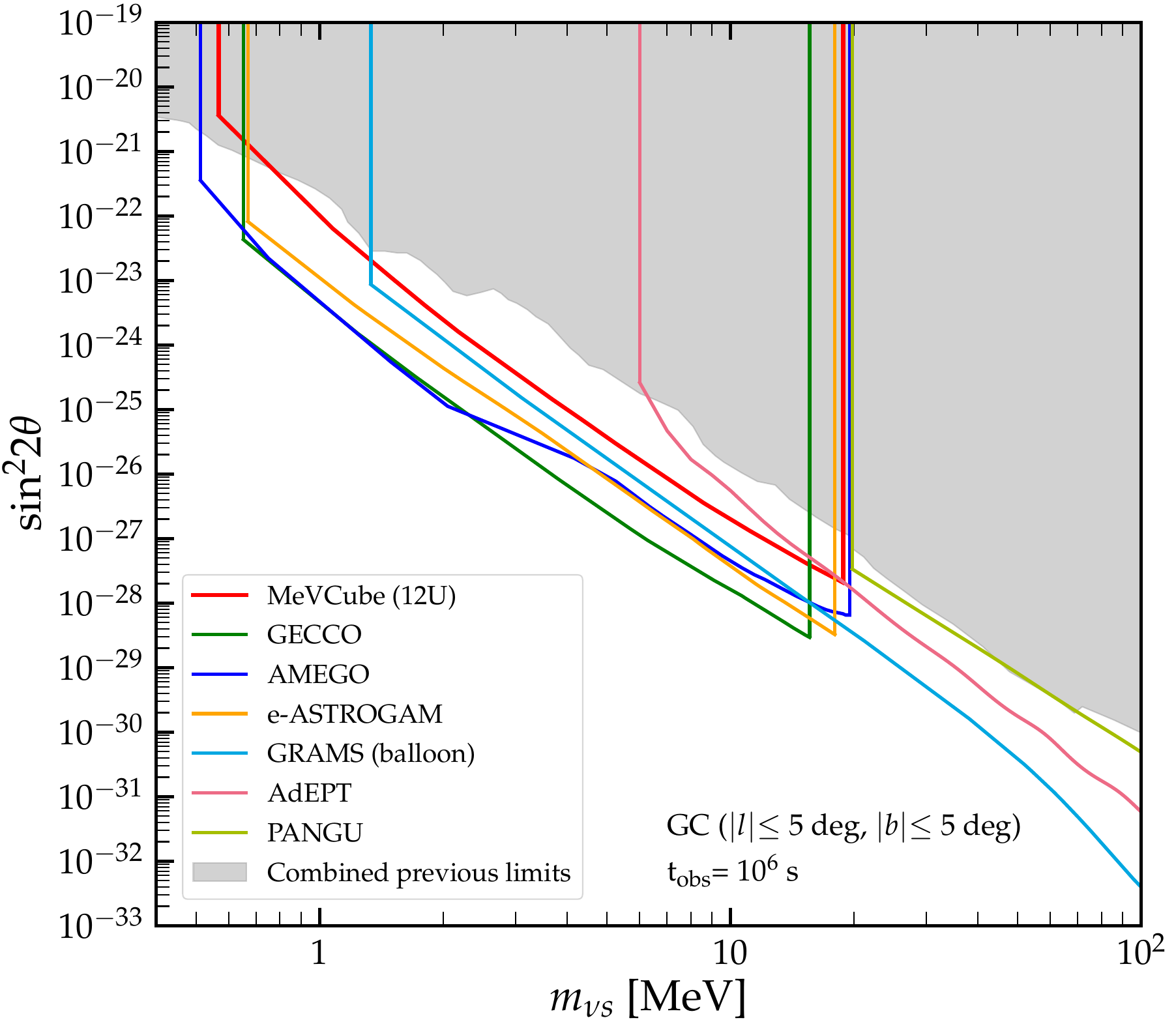}
\caption{Sensitivities of proposed MeV telescopes to sterile neutrino DM mass and mixing angle. We set observation target as the galactic centre region ($\abs{l} \leq 5$ deg, $\abs{b} \leq 5$ deg) and observation time of $10^6$\,s. The previous limits in the parameter space (grey shaded region) include inner galaxy observations from INTEGRAL\,\cite{Essig:2013goa,Calore:2022pks}, COMPTEL\,\cite{Essig:2013goa}, an earlier analysis with INTEGRAL datasets\,\cite{Boyarsky:2007ge} and NuSTAR observations\,\cite{Roach:2022lgo}.    }
\label{fig:Sterile_limits}
\end{figure}

\begin{figure*}
    \centering
    \includegraphics[width=0.65\linewidth]{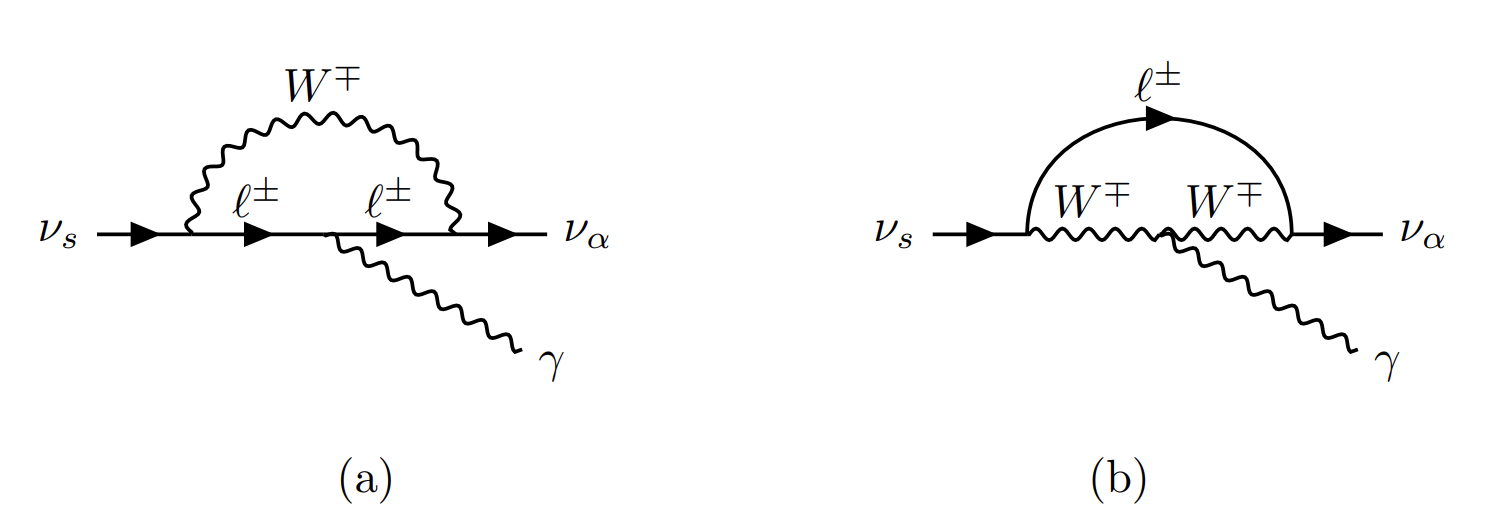}
    \centering
    \includegraphics[width=0.65\linewidth]{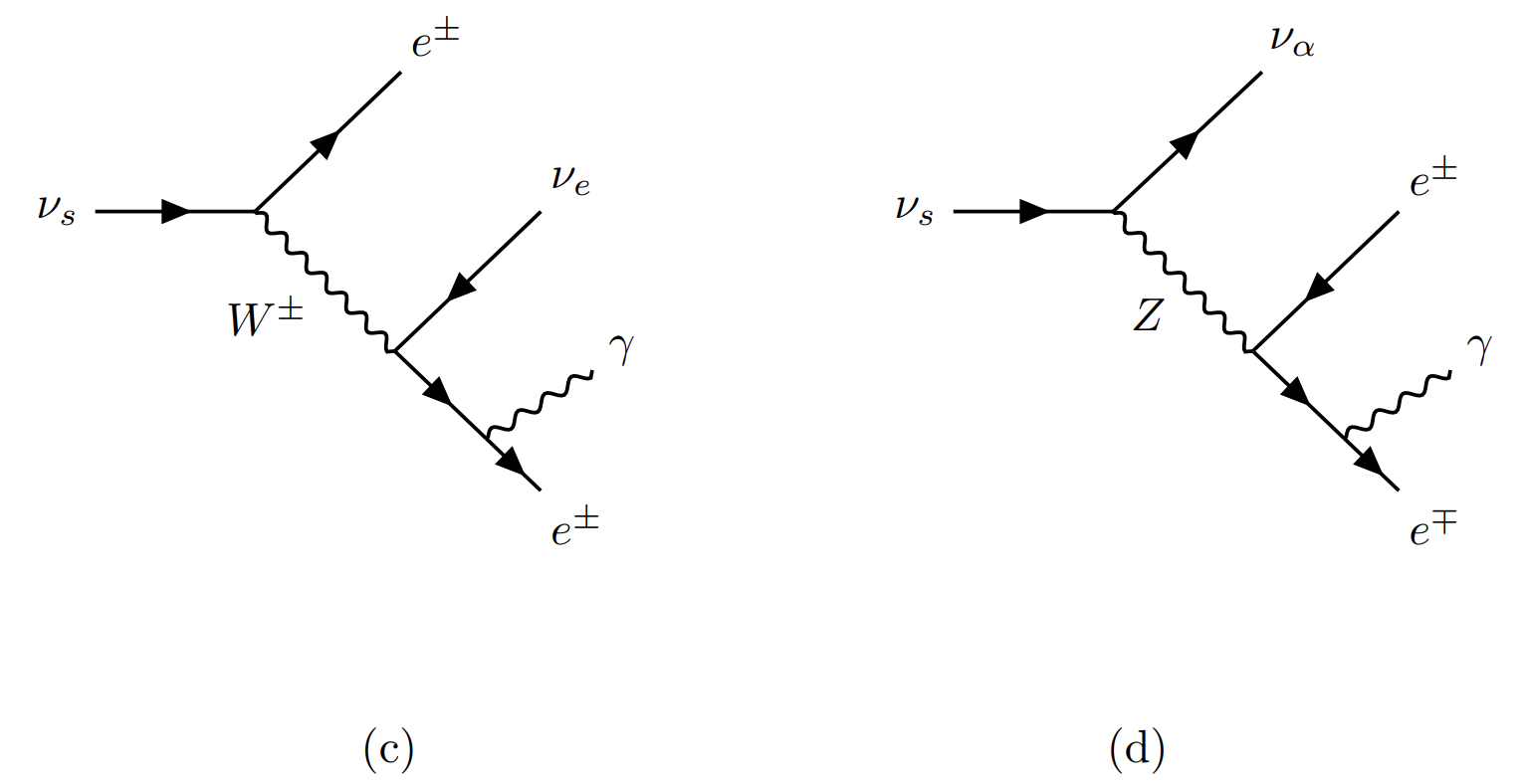}
    \caption{Feynman diagrams for the decay channels of sterile neutrino DM. \textbf{Top panel} (a,b) shows  the decay channel $\nu_s\rightarrow\nu_\alpha\,\gamma$. \textbf{Bottom panel} (c,d) shows the channel $\nu_s \rightarrow \nu_\alpha\,e^+ e^-$ + FSR.} 
    \label{fig:feynman}
\end{figure*}
One of the key strategies to discover sterile neutrino DM relies on looking for its decay products. If sterile neutrino DM decays to produce final state photons, then such excess photons can be searched for at various DM dense targets using existing telescopes. X-ray searches for such signatures have resulted in strong constraints on sterile neutrino DM in the keV mass range\,\cite{Watson:2006qb,Loewenstein:2008yi,Yuksel:2007xh,Boyarsky:2007ge, Boyarsky:2007ay, riemer2009decaying, Horiuchi:2013noa,Tamura:2014mta,Ng:2015gfa,
Neronov:2016wdd,Perez:2016tcq,Roach:2019ctw,Sicilian:2020glg,Foster:2021ngm,Krivonos:2024yvm,Calvo:2026dod}. The presence of an unidentified X-ray line signal at 3.5 keV line\,\cite{Bulbul:2014sua,Boyarsky:2014jta,Urban:2014yda,Boyarsky:2014ska,Cappelluti:2017ywp} has received significant attention since it can be explained simply by the decay of a sterile neutrino DM of mass 7.1 keV. However, the existence of the line over background is still a topic of exploration and further studies are warranted\,\cite{Horiuchi:2013noa,Malyshev:2014xqa,Jeltema:2014qfa,Anderson:2014tza,Tamura:2014mta,Jeltema:2015mee,Hitomi:2016mun,Dessert:2018qih,Dessert:2023fen}.

Significantly less attention has been devoted to the MeV regime due to historical limitations in detector sensitivity. In particular, the photon energy window of 100 keV-\,50 MeV, also known as the `MeV gap’, remains the least explored. The last dedicated telescope in this energy range was COMPTEL\,\cite{1993ApJS...86..657S}, which got decommissioned in 2000. COSI, a NASA endeavour, is planned for launch in 2027. In addition, many proposed telescopes like MeVCube\,\cite{Lucchetta:2022nrm}, the Galactic Explorer with a Coded Aperture Mask Compton Telescope (GECCO)\,\cite{Orlando_2022,Orlando:2021get}, the All-sky Medium Energy Gamma-ray Observatory (AMEGO)\,\cite{Kierans:2020otl,AMEGO:2019gny,Fleischhack:2021mhc}, the Enhanced ASTROGAM (e-ASTROGAM)\,\cite{DEANGELIS20181,e-ASTROGAM:2017pxr}, the Gamma-Ray and AntiMatter Survey (GRAMS)\,\cite{ARAMAKI2020107,Aramaki:2019bpi}, the Advanced Energetic Pair Telescope (AdEPT)\,\cite{HUNTER201418}, the PAir-productioN Gamma-ray Unit (PANGU)\,\cite{Wu:2014tya} are in discussion. These instruments provide a unique opportunity to probe sterile neutrino DM through both monochromatic photon lines and continuum emission from three-body decays. These efforts will not only revolutionise our understanding of the MeV gap but also help us uncover new physics signatures at these photon energies.

In this work, we forecast the sensitivities of the proposed MeV telescopes in searching for sterile neutrino DM through their radiative as well as three-body decays. Numerous well-motivated models exist in the literature for MeV-scale DM and the resulting production of photons from their interactions \cite{Boehm:2003hm, Pospelov:2007mp, Hochberg:2014dra, Boyarsky:2018tvu, Evans:2019jcs, Dasgupta:2021ies, Compagnin:2022elr, Chu:2022xuh, Linden:2024fby, Nguyen:2024kwy, Balan:2024cmq}. New physics searches of the proposed telescopes in the MeV gap have resulted in promising sensitivities for various DM candidates\,\cite{Boddy:2015efa,Ray:2021mxu,Coogan:2020tuf,Coogan:2021sjs,Caputo:2022dkz,Coogan:2021rez,Ghosh:2021gfa,Tseng:2022jta,Carenza:2022som,Xie:2023cwi,Berlin:2023qco,Calza:2023rjt,Kasuya:2024ldq,Cui:2024uwk,ODonnell:2024aaw,Dent:2024yje,Xie:2024eug,Agashe:2024dkq,Compagnin:2022elr,JUNO:2023vyz,Boddy:2024cyu,Manzari:2024jns,Alpine:2024kej,Buckley:2024ldr,Saha:2025wgg}. The reach of COSI in discovering MeV-scale sterile neutrino DM has been discussed recently in Ref.\,\cite{Fujisawa:2025yqi}. 
In our work, we perform Fisher forecasting analysis taking into account astrophysical background modelling to provide robust sensitivities on sterile neutrino DM using different proposed MeV satellites.  With respect to previous studies, we provide a unified treatment across multiple instruments with consistent modeling assumptions.

Our main result, shown in Fig.~\ref{fig:Sterile_limits}, depicts the sensitivity in the sterile neutrino mass-mixing parameter space. We find that when compared to existing bounds from previous X-ray observations from telescopes like INTEGRAL, COMPTEL (grey shaded region), future MeV telescopes improve constraints by several orders of magnitude.
Our work highlights the power of next-generation MeV telescopes in discovering sterile neutrino DM. 

This work is structured as follows. Section \ref{Sec:method} gives an overview of the sterile neutrino DM, the photon flux expected from its decay, and the background modeling. In section \ref{fisher} we outline the Fisher forecasting analysis, and in section \ref{Sec:instrument} a brief description of the proposed MeV telescopes. Finally, in section \ref{results} we discuss our results and future scopes.


\section{Methodology}
\label{Sec:method}

In this section, we present our methodology for calculating the gamma flux from decays of sterile neutrino DM. Considering sterile neutrinos as viable DM candidates, we proceed to calculate the signal photon flux from sterile neutrino decay. We discuss the corresponding astrophysical background model for the region of interest. These will be key ingredients for the Fisher matrix formalism that we use to derive robust sensitivities on sterile neutrino DM mass and the mixing.  
\subsection{Sterile neutrino DM}
\label{sterile}
 In this work, we remain agnostic of the sterile neutrino DM production mechanism and focus on the observational signatures arising from decay in the MeV gap. Regardless of the production mechanism, a sterile neutrino, due to its mixing with an active neutrino, can radiatively decay ($\nu_s\rightarrow\nu_\alpha\,\gamma$)  to produce a final state neutrino and a photon, where \(\nu_\alpha\) is one of the active neutrino states (Feynman diagrams shown in Fig.\,\ref{fig:feynman} (a,b)). Since sterile neutrinos are significantly heavier than active neutrinos, the emitted monochromatic photon in this two-body decay has an energy, \(E_{\gamma}=m_{\nu_s}/2\).

The decay width for this channel is given as\,\cite{PhysRevD.9.743,Pal:1981rm,Bezrukov:2009th}
\begin{align}
\Gamma_{\nu\gamma} &\simeq \frac{9\,\alpha_{\rm EM}\, G_F^2\,m_{\nu_s}^5\sin^2(2\theta)}{1024\,\pi^4} \nonumber \\
& \simeq 1.39 \times 10^{-18} ~ {\rm s}^{-1}
\left(\frac{m_{\nu_s}}{10~ {\rm MeV}} \right)^5\left[\frac{ \sin^2(2\theta)}{10^{-16}} \right]
\label{eq:nuwidth}
\end{align}
where $\theta$ is the mixing angle between the sterile and active neutrino and $\alpha_{\rm EM}=1/137$. 

For \(m_{\nu_s} > 2m_e \gtrsim 1\) MeV, yet another decay channel opens up where sterile neutrino decays to an active neutrino and an electron-positron pair (\(\nu_s \rightarrow \nu_\alpha\,e^+ e^-\)). These final-state electron-positron pairs can also produce continuum gamma rays via final-state radiation (FSR) (Feynman diagrams shown in Fig.\,\ref{fig:feynman} (c,d)). The decay rate for this process is given by\,\cite{Ruchayskiy:2011aa}
\begin{align}
\Gamma_{\nu_\alpha e^+ e^-}
&\simeq 
\frac{c_\alpha \sin^2(2\theta)}{384\,\pi^3}
G_F^2\,m_{\nu_s}^5 \nonumber \\
&\simeq
1.04 \times 10^{-16}\,\text{s}^{-1}
\left(\frac{m_{\nu_s}}{10\,\text{MeV}}\right)^5
\left[\frac{\sin^2(2\theta)}{10^{-16}}\right]
\end{align}
with the prefactor depending on the neutrino flavor and weak couplings. In the above equation, we consider the neutrino flavor $\alpha=e$ and $c_\alpha \simeq 0.59$.

The FSR photon spectrum is broad and subdominant compared to the monochromatic line in most of the parameter space considered here, but is included for completeness. See Fig.\,\ref{fig:FSR comparison} for a comparison in sensitivities for both the channels (discussed in Sec.\,\ref{results}).

\subsection{Signal modeling}
\label{signal}
The photon flux produced from the decay of the sterile neutrino is given by 
\begin{align}
\phi_{\rm DM}(E_\gamma) = \frac{r_\odot}{4\pi} \frac{\rho_\odot}{m_{\nu_s} } \left( \frac{d^2 N}{dE_\gamma dt} \right) \frac{D}{\Delta \Omega}\,\,,    
\label{fluxdm}
\end{align}
where $\phi_{\rm DM}(E_\gamma)$ has units of MeV$^{-1}$cm$^{-2}$s$^{-1}$sr$^{-1}$. In the above equation, \(\rho_\odot = 0.4 \, \text{GeV/cm}^3\) is the DM density in the solar neighborhood, \(r_\odot = 8.3 \, \text{kpc}\) is the distance between the Solar System and the center of the Milky Way (MW), \(\left( \frac{d^2 N}{dE_\gamma dt} \right)\) is the spectrum of photons produced per DM decay per unit time, \(\Delta \Omega\) is the angular size of the region of interest, and \(D\) is the D-factor that incorporates the DM distribution defined as:
\begin{align}
D = \int \frac{ds}{r_\odot} \left( \frac{\rho}{\rho_\odot} \right) d\Omega    
\end{align}
where \(\rho\) is the DM density profile within the MW and \(s\) is the line-of-sight distance. The density profile of DM can be parameterized as \cite{Bringmann:2012ez}:
\begin{align}
\rho_\chi^{abc}(r) = \rho_\odot \left( \frac{r}{r_\odot} \right)^{-c} \left[ \frac{1 + (r_\odot / r_s)^a}{1 + (r / r_s)^a} \right]^{(b - c) / a}  
\label{dmprof}
\end{align}
where \(r\) is the distance from our Galactic Center (GC), and \(r_s\) is the scale radius. The distance \(r\) and line-of-sight distance \(s\) are related by:
\begin{align}
r(s, b, \ell) = \sqrt{s^2 + r_\odot^2 - 2 s r_\odot \cos b \cos \ell}    
\end{align}
where \(l\) and \(b\) are the galactic longitude and latitude of the region of interest, respectively. The differential angular size is \(d\Omega = d(\sin b) \, d\ell\).

In this work, as a benchmark, we use the widely-used Navarro–Frenk–White (NFW) profile as the DM density profile\,\cite{Navarro:1996gj} and choose \((a, b, c) = (1, 3, 1)\) and \(r_s = 20 \, \text{kpc}\) in Eq.\,(\ref{dmprof}). 

For the  $\nu_s\rightarrow\nu_\alpha\,\gamma$ channel, the resulting photon spectra is given by\,\cite{Essig:2013goa} 
\begin{align}
 \frac{d^2N}{dE_\gamma dt}=\Gamma_{\nu\gamma}\times\delta\left(E_\gamma-\frac{m_{\nu_s}}{2}\right)\,.
 \label{eq:d2ndedt}
\end{align}

In the case of \(\nu_s \rightarrow \nu_\alpha\,e^+ e^-\) + FSR channel, the spectra is\,\cite{Essig:2013goa} 
\begin{align}
\frac{d^2N_{\text{FSR}}}{dE_\gamma dt} \simeq 
&\;\Gamma_{\nu_\alpha e^+ e^-}\times\frac{2\alpha_{\rm EM}}{\pi E_\gamma}
\log\left(\frac{1 - 2\lambda_\gamma}{\xi_e^2}\right) \nonumber \\
&\times \Bigg[
1 - \frac{11}{3}\lambda_\gamma + 10\lambda_\gamma^2 
+ \frac{\lambda_\gamma (1 + 4\sin^2\theta_W)(1 - 6\lambda_\gamma)}{12c_\alpha} \nonumber \\
& + \text{Higher order term in $\xi_e$}
\Bigg] 
\end{align}
where  $\lambda_\gamma=E_\gamma/m_{\nu_s}$, $\xi_e=m_e/m_{\nu_s}$, and $\sin^2\theta_W\sim0.23$.

\subsection{Background modeling}
\label{bkg}

To understand the prospect of upcoming MeV telescopes in discovering sterile neutrino decay, it is essential to carefully model the known astrophysical backgrounds in the energy range of interest. In this work, we focus on the Milky Way galactic centre region \((|l|\leq5^\circ, |b|\leq5^\circ)\).  The dominant backgrounds in the MeV range arise from galactic and extragalactic gamma-ray emission, and can be parametrized following Refs.\, \cite{Bartels:2017dpb,Beacom:2005qv,Ray:2021mxu}. For the galactic background, we consider 3 components -- an inverse Compton scattering (ICS) component that is fit to the COMPTEL dataset, a higher energy ICS component fit to the Fermi-LAT data, and a $\pi^0$ component. The low energy ICS component can be parametrized as,
 \begin{align}
     \phi_{\rm ICS_{\rm lo}}^{\rm bkg}(E_\gamma)= A_{\rm g}^{\rm bkg}\left(\frac{E_\gamma}{1\,\rm MeV}\right)^{-\alpha^{\rm g}}{\rm exp}\left[-\left(\frac{E_\gamma}{E_c}\right)^{\bar{\gamma}}\right]\,\,
     \label{galbkg}
 \end{align}
 where $A_{\rm g}^{\rm bkg}$ is the normalization, $\alpha^{\rm g}$ is the spectral index, $\bar{\gamma}$ is the index within
the exponential, and $E_c$ is the exponential cut-off energy.
 For the higher energy ICS component and the $\pi^0$ component, we use the flux given in Ref.\,\cite{Bartels:2017dpb} and assume overall normalization parameters $A_{\rm ICS_{\rm hi}}$ and $A_{\rm PI}$ as free parameters. 
 
 The extragalactic background component can also be modeled as a power law,
 \begin{align}
     \phi_{\rm eg}^{\rm bkg}(E_\gamma)=A_{\rm eg}^{\rm bkg}\left(\frac{E_\gamma}{1\,\rm MeV}\right)^{-\alpha^{\rm eg}}\,\,
      \label{exgalbkg}
\end{align}
with $A_{\rm eg}^{\rm bkg}$ and $\alpha^{\rm eg}$ being the normalization and spectral index, respectively.

The backgrounds are usually quoted in MeV$^{-1}$cm$^{-2}$s$^{-1}$sr$^{-1}$. The fiducial background parameter values for the above backgrounds are, $A_{\rm g}^{\rm bkg}= 0.013$ MeV$^{-1}$cm$^{-2}$s$^{-1}$sr$^{-1}$, $\alpha^{\rm g}= 1.8$, $\bar{\gamma}=2$, $E_c=20$ MeV, $A_{\rm eg}^{\rm bkg}=0.004135$ MeV$^{-1}$cm$^{-2}$s$^{-1}$sr$^{-1}$, $\alpha^{\rm eg}=2.8956$, $A_{\rm ICS_{\rm hi}}=1$, and $A_{\rm PI}=1$. We note that in our analysis we marginalize over these parameters to obtain the final sensitivities on sterile neutrino DM mass and mixing. We discuss this in detail in the next section. 

To evaluate the measured photon flux from sterile neutrino DM, we have to take into account the finite energy resolution of the respective telescopes. The measured photon flux is given by~\cite{Bringmann:2008kj}
\begin{align}
\left(\frac{d^2N}{dE_\gamma dt}\right)(E_\gamma) = \int dE_\gamma^\prime \,\mathcal{R}_\epsilon(E_\gamma |E_\gamma^\prime)\left(\frac{d^2N}{dE_\gamma dt}\right)^\prime(E_\gamma^\prime)    
\end{align}
where
\begin{eqnarray}
    \mathcal{R}_\epsilon(E_\gamma|E_\gamma^\prime) = \frac{1}{\sqrt{2 \pi(\epsilon E_\gamma)^2}}{\rm exp}\left[\frac{-(E_\gamma-E_\gamma^\prime)^2}{2 (\epsilon E_\gamma)^2}\right]
\end{eqnarray}
is the finite energy resolution for a given telescope, modeled as a Gaussian function. This quantity specifies the probability that a photon with true energy \(E_\gamma^\prime\) will be detected as a photon with energy \(E_\gamma\). The energy resolutions for each of the proposed telescopes are provided in the respective references (provided in the next section).  

Indeed, for a line search like $\nu_s\rightarrow\nu_\alpha\,\gamma$ channel, the energy resolution of a telescope becomes very important since it determines the finite width of the line. On the other hand, for a broad spectrum like the \(\nu_s \rightarrow \nu_\alpha e^+ e^-\) + FSR channel, the energy resolution has minimal effect. The total photon flux seen by a telescope then consists of both the sterile DM-origin flux and the background flux:

\begin{align}
\Phi(\theta) = \phi_{\rm DM}(\theta_{\chi}) + \phi^{\rm bkg}(\theta_{\rm bkg})\,\,    
\end{align}
where \(\theta_{\chi}\) and \(\theta_{\rm bkg}\) are the sterile DM model parameters and astrophysical background parameters, respectively. For decaying DM, the model parameters are $\theta_{\chi} = (\Gamma_{\nu\gamma}, \Gamma_{\nu_\alpha e^+ e^-})$. We evaluate the sensitivities for each of the sterile neutrino decay channels individually and then combine the limits.

The background flux takes into account both galactic and extragalactic components:

\begin{align}
\phi^{\rm bkg}(\theta_{\rm bkg}) = \phi_{\rm g}^{\rm bkg}(\theta_{\rm bkg}) + \phi_{\rm eg}^{\rm bkg}(\theta_{\rm bkg})    
\end{align}
where $\theta_{\rm bkg} = (A_{\rm g}^{\rm bkg}, \alpha^{\rm g}, \bar{\gamma}, E_c, A_{\rm eg}^{\rm bkg}, \alpha^{\rm eg}, A_{\rm ICS_{\rm hi}}, A_{\rm PI})$ denotes the parameters of the background model. 

\section{Fisher forecasting analysis}
\label{fisher}

We use the Fisher forecasting method \cite{Edwards:2017mnf} and marginalize over all background parameters to estimate the sensitivity of future MeV telescopes to sterile neutrino DM. This approach provides an estimate of the expected parameter uncertainties under the assumption of a Gaussian likelihood and sufficiently large event counts. The Fisher matrix elements are defined as

\begin{align}
    \mathcal{F}_{ij} = \int dE_\gamma\,d\Omega\,t_{\rm obs}\,A_{\rm eff}(E_\gamma) \left[\frac{1}{\Phi}\left(\frac{\partial\Phi}{\partial \theta_i}\right)\left(\frac{\partial\Phi}{\partial \theta_j}\right)\right]_{\theta=\theta_{\rm fid}}\,,
\end{align}
where $t_{\rm obs} =10^6$ seconds is the observation time, \( A_{\rm eff}(E_\gamma) \) denotes the effective area of the respective telescopes taken from the corresponding references (given in the next section), and \( \theta_{\rm fid} \) specifies the fiducial value for a parameter \( \theta \). For the sterile neutrino DM signal, we assume a background-only fiducial model, \( (\theta_\chi)_{\rm fid}=0 \), which corresponds to the null hypothesis of no dark matter signal and allows us to compute projected upper limits.

For telescopes with energy thresholds below 10 MeV, we include only the low-energy ICS component, $\phi_{\rm ICS_{\rm lo}}^{\rm bkg}$, as the galactic background component, given that the other components are subdominant (see Fig.\,2 of Ref.\,\cite{Bartels_2017}). In these cases, the Fisher matrix \( \mathcal{F} \) is a 7 × 7 symmetric matrix, which includes one model parameter, $\Gamma_{\nu\gamma}$ or $\Gamma_{\nu_\alpha e^+ e^-}$, and six background parameters (\( A_{\rm g}^{\rm bkg}, \alpha^{\rm g}, \bar{\gamma}, E_c, A_{\rm eg}^{\rm bkg}, \alpha^{\rm eg} \)). Assuming that a given telescope will not detect any DM signal during its observation time, we can establish a 95\% confidence level upper limit on the DM model parameter \( \theta_{\chi} \) as follows \cite{Edwards:2017mnf}:

\begin{align}
    \theta_{\chi}^{\rm UL} = 1.645 \sqrt{(\mathcal{F}^{-1})_{11}}\,.
\end{align}
Here, \( \theta_{\chi} \) corresponds to $\Gamma_{\nu\gamma}$ or $\Gamma_{\nu_\alpha e^+ e^-}$. 
For MeV telescopes like PANGU and AdEPT, which are sensitive to higher energy photons, we take into account the higher energy ICS and $\pi^0$ components, and the Fisher matrix \( \mathcal{F} \) becomes a 9 × 9 symmetric matrix.  

The matrix $\mathcal{F}$ also contains the correlations between the signal and background parameters. Thus, given a Fisher matrix, one can estimate the uncertainties in different parameters along with the correlations between them. The correlation coefficient between $i^{\rm th}$ and $j^{\rm th}$ parameter is defined as
\begin{eqnarray}
    \rho_{ij}=\frac{C_{ij}}{\sqrt{C_{ii}\times C_{jj}}}\,\,
    \label{correlation}
\end{eqnarray}
where $C_{ij}$ is the covariance matrix element that is related to the Fisher matrix element by
\begin{eqnarray}
    C_{ij}= \mathcal{F}_{ij}^{-1}.
\end{eqnarray}
the correlation coefficient, $\rho_{ij} \in [-1,1]$, where $\rho_{ij}>0$, $\rho_{ij}<0$, and $\rho_{ij}=0$ means positive, negative, and zero correlations between $i^{\rm th}$ and $j^{\rm th}$ parameters, respectively. 


\section{Instrument details}
\label{Sec:instrument}
 In this section, we outline the proposed MeV telescopes that we consider for our work and their specifications. The launch timelines of these telescopes are not clear yet.  Given their different, complementary energy ranges of interest and sensitivities, we choose to evaluate the capabilities of all of these telescopes in searching for sterile neutrino DM. We note that the list of telescopes given here is not complete, and there are other proposed telescopes under consideration. As a benchmark, we only focus on the following telescopes for the search for sterile neutrino DM.

\subsection{MeVCube}
\label{MevCube}
MeVCube is a proposed CubeSat-standard telescope designed to explore the MeV gap~\cite{Lucchetta:2022nrm,Lucchetta:2022zrd}. It will consist of two layers of pixelated Cadmium-Zinc-Telluride (CdZnTe) spaced 6 cm apart, with the first layer for scattering and the second for absorption. The final size of MeVCube is yet to be determined. We consider a 12U configuration in this work. MeVCube will be sensitive to photon energies between 200 keV - 4 MeV with a wide field of view ($\sim$ 2 sr) and angular resolution of 1.5° at 1 MeV. The specifications of MeVCube are taken from Ref.\,\cite{Lucchetta:2022nrm}.

\subsection{GECCO}
\label{Gecco}
The Galactic Explorer with a Coded Aperture Mask Compton Telescope (GECCO) combines a Compton telescope with a coded mask telescope in the energy range of 0.2 to 8 MeV. The expected energy resolution is less than 1\% for energies between 0.5 and 5 MeV, while the expected angular resolution is approximately 1 arc minute for the 4 × 4 degree coded aperture mask. In contrast, the angular resolution for the analysis with the 60 × 60 degree Compton telescope data ranges from 4 to 8 degrees. An overview of GECCO is presented in \cite{Orlando_2022}.

\subsection{AMEGO}
\label{amego}
The All-sky Medium Energy Gamma-ray Observatory (AMEGO) is a proposed telescope that can combine two different detection modes: Compton scattering and pair production. As a result, the energy range of interest is wide ($\sim 200$ keV -- 5 GeV).
Energy resolution for the Compton mode is provided in \cite{Kierans:2020otl}, while we assume a conservative 30\% resolution for the pair production mode. A detailed  overview of AMEGO is available in Ref.\,\cite{Kierans:2020otl}. A recent proposal, AMEGO-X \cite{Caputo_2022}, offers comparable effective area and energy resolution, suggesting similar sensitivity. In this work we use AMEGO sensitivities due to detailed available specifications.

\subsection{e-ASTROGAM}
\label{eastrogam}
The Enhanced ASTROGAM (e-ASTROGAM) also uses two modes for exploring the MeV Gap: a Compton scattering and a pair-production mode. The Compton scattering mode has $\sim$ 1\% energy resolution, while the pair-production mode has $\sim$ 20\%--30\% energy resolution. For our work, we assume an energy resolution of 30\% ($\Delta E / E$) throughout the pair-production energy range. An overview of e-ASTROGAM is given by \cite{DEANGELIS20181}.

\subsection{GRAMS}
\label{grams}

The Gamma-Ray and AntiMatter Survey (GRAMS) employs a liquid argon time-projection chamber and can operate within the 100 keV to 100 MeV energy range, with an estimated energy resolution of about 1\%, coarser at lower energies \cite{ARAMAKI2020107}. At higher energies, its pair-production mode increases the effective area but reduces the energy resolution. An overview of GRAMS is provided in \cite{Aramaki:2021o5}. Recently, a balloon flight with a small-scale setup was conducted \cite{Nakajima:2024fgx}. This is an important step towards a planned satellite version with upgraded detectors. An overview of GRAMS can be found in \cite{ARAMAKI2020107}.

\subsection{AdEPT}
\label{adept}
The Advanced Energetic Pair Telescope (AdEPT) is a pair-production telescope sensitive within the energy range of 5 to 200 MeV. The effective area for AdEPT is given in \cite{HUNTER201418}, which also reports an energy resolution of 30\% FWHM at 70 MeV. For our work, we assume that this energy resolution remains consistent throughout the entire energy range of the instrument. An overview of AdEPT can be found in \cite{HUNTER201418}.

\subsection{PANGU}
\label{pangu}
The PAir-productioN Gamma-ray Unit (PANGU) is a pair-production telescope that will operate within the energy range of 10 MeV to 1 GeV. Ref.\,\cite{Wu:2014tya} provides an upper limit on the energy resolution of 50\% $(\Delta E / E)$, and we utilized this value across its entire energy range for our calculations. An overview of PANGU can be found in \cite{Wu:2014tya}.


\section{Results and Discussions}
\label{results}

In Fig.~\ref{fig:Sterile_limits}, we present the projected sensitivities of several proposed MeV telescopes to sterile neutrino DM. Existing constraints in the relevant parameter space include limits from diffuse Galactic gamma-ray emission observed by INTEGRAL~\cite{Essig:2013goa,Calore:2022pks} and COMPTEL~\cite{Essig:2013goa}, as well as combined limits from X-ray observations~\cite{Boyarsky:2007ge,Roach:2022lgo}, shown as the gray shaded region. We find that the projected sensitivities of the proposed MeV telescopes improve upon the strongest existing limits by several orders of magnitude over a broad range of sterile neutrino masses. Owing to their sensitivity to higher-energy photons, telescopes such as AdEPT, PANGU, and GRAMS can probe heavier sterile neutrino DM than the other proposed instruments. 
Since existing constraints from Big Bang nucleosynthesis (BBN)~\cite{Serpico:2005bc} and subhalo counts~\cite{Cherry:2017dwu} are model dependent, we do not include them in Fig.~\ref{fig:Sterile_limits}.

\begin{figure}[t]
\centering
\includegraphics[width=\columnwidth]{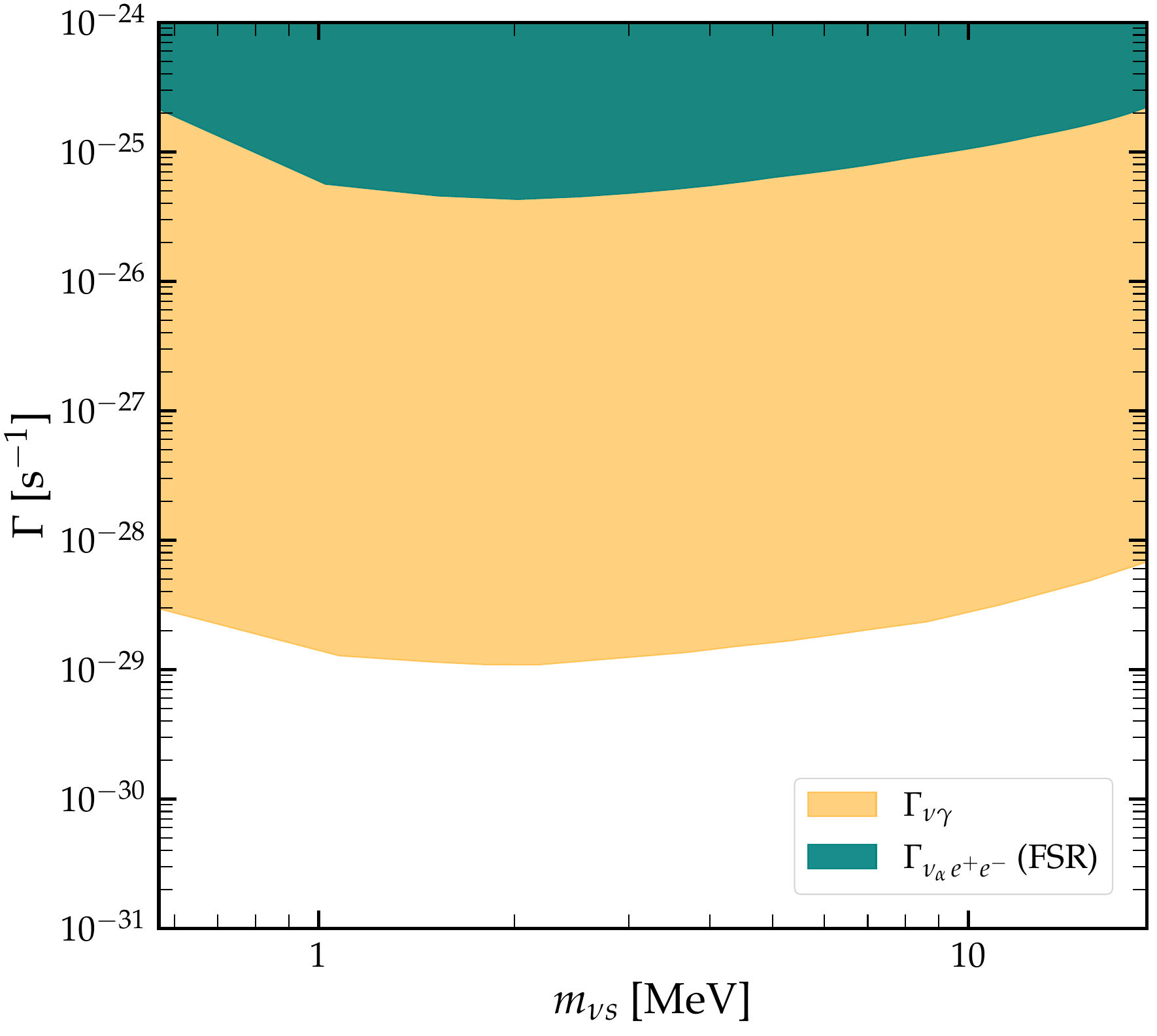}
\caption{ Sensitivities of the MeVCube (12U) telescope to sterile neutrino DM decay rates as a function of its mass for the $\nu_s\rightarrow\nu_\alpha\,\gamma$ channel (yellow shaded region) and $\nu_s \rightarrow \nu_\alpha e^+ e^-$ + FSR channel (teal shaded region). For both of these cases, we use the galactic centre region ($\abs{l} \leq 5$ deg, $\abs{b} \leq 5$ deg) as the target and an observation time of $10^6$\,s.}
\label{fig:FSR comparison}
\end{figure}

The sensitivities presented in this work correspond to the combined best limits obtained from the radiative decay channel, $\nu_s \rightarrow \nu_\alpha \gamma$, and the three-body decay channel with final-state radiation (FSR), $\nu_s \rightarrow \nu_\alpha e^+ e^-$. We find that the radiative decay channel dominates the sensitivity across the full sterile neutrino mass range considered in this work. In Fig.~\ref{fig:FSR comparison}, we compare the projected sensitivities to the decay rate for the $\nu_s \rightarrow \nu_\alpha \gamma$ channel (yellow shaded region) and the $\nu_s \rightarrow \nu_\alpha e^+ e^-$ (FSR) channel (teal shaded region), taking MeVCube as a representative example. For this comparison, we assume the same target region and observation time as in Fig.~\ref{fig:Sterile_limits}. Our results are consistent with the findings of Ref.~\cite{Calore:2022pks}. Although our analysis can, in principle, be extended beyond 100 MeV, constraints from Fermi-LAT are expected to become dominant at sufficiently high masses. We therefore leave a dedicated analysis using Fermi-LAT data for future work.

\begin{figure}
    \centering
    \includegraphics[width=1.05\textwidth, trim=0 0 0 0, clip]{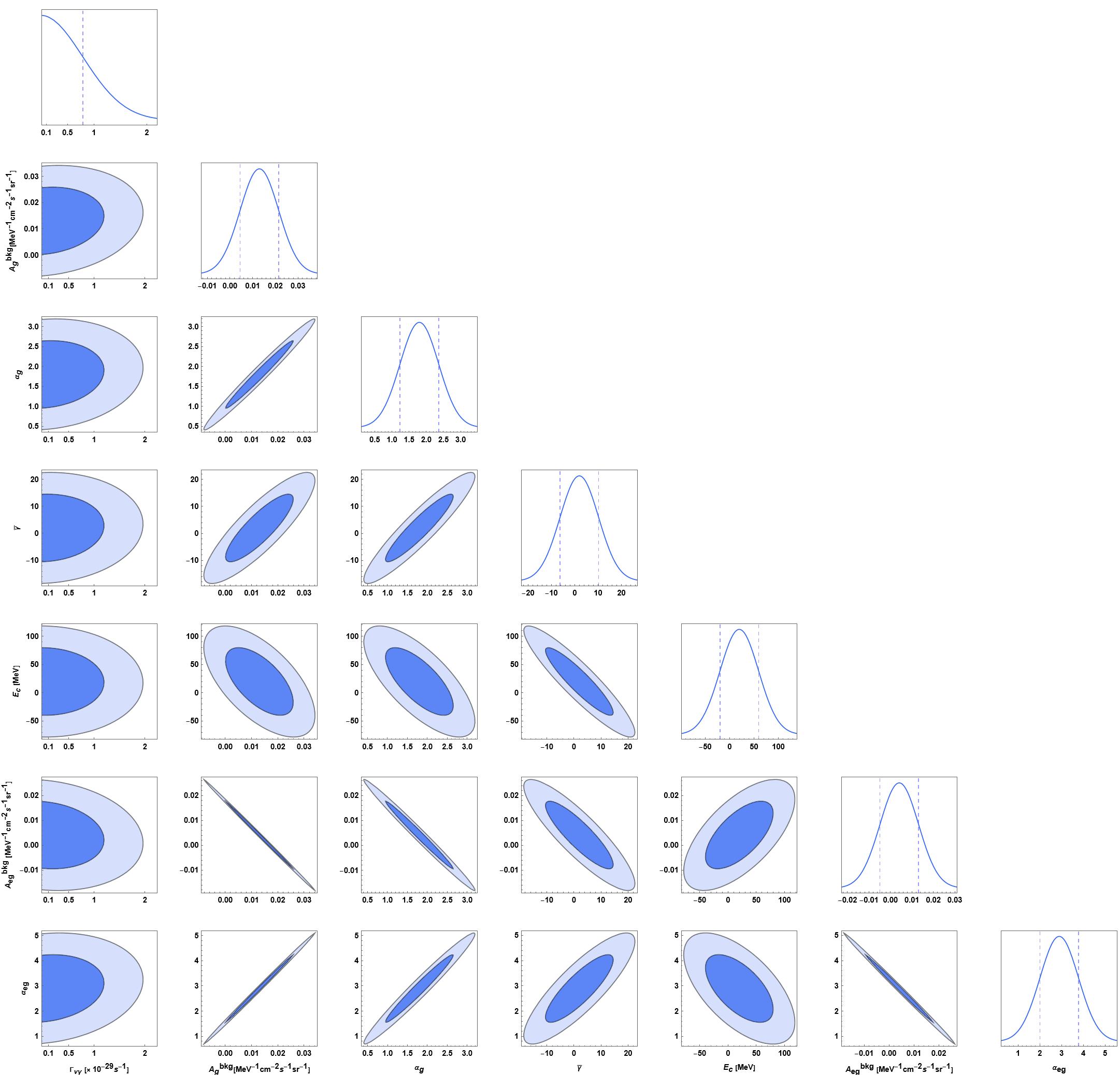}
    \begin{minipage}{\textwidth}
    \caption{Fisher forecast constraints on the model and the background parameters from MeVCube (12U). We consider $m_{\nu_s}=1$ MeV and $t_{\text{obs}}=10^6$s. The dashed lines in distributions correspond to $1\sigma$, whereas in the contour plot, the dark and light shaded regions correspond to $1\sigma$ and $2\sigma$, respectively. For the posterior of the decay rate, we only show positive values as negative decay rate is not physical.}
    \label{fig:correlations}
    \end{minipage}
\end{figure}

In Fig.~\ref{fig:correlations}, we show the projected uncertainties and correlations between the signal and background parameters for MeVCube. As a benchmark, we consider a sterile neutrino DM mass of 1 MeV and an observation time of $10^6$~s. The model parameter shown is the decay rate for the radiative channel, $\nu_s \rightarrow \nu_\alpha \gamma$. The dark and light blue shaded ellipses represent the projected $1\sigma$ and $2\sigma$ confidence regions, respectively. These contours illustrate the parameter degeneracies in the analysis. For example, the correlation between the Galactic background normalization, $A_{\rm g}^{\rm bkg}$, and the spectral index, $\alpha^{\rm g}$, indicates that an increase in $A_{\rm g}^{\rm bkg}$ can be partially compensated by an increase in $\alpha^{\rm g}$ while maintaining a comparable fit to the data. This implies a positive correlation between these two parameters. We also show the marginalized one-dimensional posteriors for each parameter.

The correlation coefficient between the sterile neutrino decay rate and the other background parameters are: $\rho(\Gamma_{\nu\gamma},A_{\rm g}^{\rm bkg})=0.15$, $\rho(\Gamma_{\nu\gamma},\alpha^{\rm g})=-0.12$, $\rho(\Gamma_{\nu\gamma},\bar{\gamma})=-0.07$, $\rho(\Gamma_{\nu\gamma},E_c)=0.03$, $\rho(\Gamma_{\nu\gamma},A_{\rm eg}^{\rm bkg})=0.15$, and ~$\rho(\Gamma_{\nu\gamma},\alpha^{\rm eg})=-0.16$.
The weak correlations between the sterile neutrino decay rate and the background parameters, with $|\rho| \lesssim 0.15$, indicate that the DM signal is spectrally distinct from the dominant astrophysical backgrounds. This demonstrates that the projected sensitivities are not significantly affected by parameter degeneracies and remain robust under marginalization. At the same time, some background parameter pairs, such as $A_{\rm eg}^{\rm bkg}$--$A_{\rm g}^{\rm bkg}$ and $\alpha^{\rm eg}$--$A_{\rm g}^{\rm bkg}$, exhibit narrow elliptical contours, indicating stronger correlations. This implies that these background parameters cannot be independently constrained with the same level of precision.

The results presented in this work assume observations of the Galactic center region by the proposed MeV telescopes. Dwarf spheroidal galaxies provide another promising target for DM searches because of their high DM densities and relatively low astrophysical backgrounds. However, sensitivity studies for decaying DM in dwarf galaxies are typically weaker than those for the Galactic center~\cite{Caputo:2022dkz,ODonnell:2024aaw}. Nevertheless, future observations of dwarf galaxies will provide an important complementary probe for DM discovery.

In addition to the astrophysical background, instrumental systematics also play a crucial role in determining the DM discovery potential of the upcoming MeV telescopes. Each collaboration performs detailed in-lab measurements and simulations to understand their systematic noises. Given the lack of public information about the systematics of the upcoming MeV telescopes, we do not take them into account for our analysis. As more detailed detector performance information becomes available, our sensitivity projections can be further refined.

The electron-positrons produced from the channel $\nu_s\rightarrow \nu_\alpha\,e^+\,e^-$, can lose energy and form a positronium bound state, which eventually can produce 511 keV line emission. Given the observations of 511 keV galactic centre excess\,\cite{Weidenspointner:2004my,Churazov:2004as,Jean:2005af,Weidenspointner:2007rs,Weidenspointner:2008zz}, the implications for sterile neutrino DM can be very interesting. A dedicated study of this signal would require a detailed treatment of charged-particle propagation and energy loss in the Galactic environment. We leave such an analysis for future work.

Given the production of final state neutrinos in all the decay channels for sterile neutrino DM, detection of these MeV-scale neutrinos may be possible. The neutrinos from sterile neutrino DM decay will contribute to the diffuse supernova neutrino background (DSNB)\,\cite{Ando:2004hc,Beacom:2010kk,Lunardini:2010ab}. In principle, one can thus obtain bounds from Super-Kamiokande upper limits on the DSNB\,\cite{Super-Kamiokande:2011lwo} or even sensitivities for the upcoming Hyper-Kamiokande. Given the weak nature of these limits, we do not show them in our results. With future improvements in the detection frontier, this channel will become a very strong observational probe for sterile neutrino DM.

Our sensitivity projections assume an observation time of $10^6$~s, corresponding to approximately 12 days. This is a conservative choice, and longer observation time will improve our sensitivities as $\sqrt{t_{\rm obs}}$. Since our region of interest is centered on the Galactic center, the projected sensitivities also depend on the assumed DM density profile. Relative to the NFW profile adopted in this work, more cuspy profiles would strengthen the projected limits, while cored profiles would weaken them. Given the large number of proposed MeV-gap telescopes, our results demonstrate the strong potential of upcoming observations to probe previously unexplored regions of sterile neutrino DM parameter space. With continued advances in both detector technology and theoretical modeling, the MeV gamma-ray window may provide a powerful avenue toward identifying the nature of DM.

\section*{Acknowledgments}
We thank Anirban Das and Ranjan Laha for detailed discussions. AKS acknowledges the Ministry of Human Resource Development, Government of India, for financial support via the Prime Minister’s Research Fellowship (PMRF). MS acknowledges support from the Early Career Research Grant by Anusandhan National Research Foundation (project number ANRF/ECRG/2024/000522/PMS). MS also acknowledges support from the IoE-funded Seed Funding for Collaboration and Partnership Projects - Phase IV SCPP grant (RD/0524-IOE00I0-012) by IIT Bombay.


\newpage 
\clearpage


\newpage

\bibliographystyle{JHEP}
\bibliography{ref.bib}

\end{document}